\documentstyle[12pt]{article}

\input{tcilatex}

\begin{document}

\title{On the quasi - exact solvability of a singular potential in D - dimensions:
confined and unconfined}
\author{Omar Mustafa \\
Department of Physics, Eastern Mediterranean University\\
G. Magusa, North Cyprus, Mersin 10 - Turkey\\
email: omar.mustafa@emu.edu.tr\\
}
\maketitle

\begin{abstract}
{\small The D - dimensional quasi - exact solutions for the singular even -
power anharmonic potential $V(q)=a q^2+b q^{-4}+cq^{-6}$ are reported. We
show that whilst Dong and Ma's [5] quasi - exact ground - state solution (in
D=2) is beyond doubt, their solution for the first excited state is exotic.
Quasi - exact solutions for the ground and first excited states are also
given for the above potential confined to an impenetrable cylindrical ( D=2
) or spherical ( D=3 ) wall.}
\end{abstract}

\newpage

It is well known that a straightforward generalization of an elementary
ansatz, for a particular state wave function, defines the related quasi
(partially) - exact solvability of \newline
\begin{equation}
V(q)=aq^{2}+bq^{-4}+cq^{-6}~;~a~,c>0.
\end{equation}
\newline
A model which is phenomenologically appealing for physicist and
mathematicians alike [1-6, and references therein]. The study of such a
singular potential model ( a non - trivial generalization of the simple
spiked harmonic oscillator from the mathematical point of view) has
relevance in connection with the imaginary time formation of quantum
mechanics and its relation with diffusion theory [3]. It has been widely
used in atomic and optical physics [6].

On the other hand, the underlying time - independent radial Schr\"{o}dinger
equation, in $\hbar =2m=1$ units, \newline
\begin{equation}
\left[ -\frac{d^{2}}{dq^{2}}+\frac{L(L+1)}{q^{2}}+V(q)-E_{k,L}\right]
R_{k,L}(q)=0,
\end{equation}
\newline
admits interdimensional degeneracies associated with the isomorphism between
orbital angular momentum $l$ and dimensionality $D$ in $L=l+(D-3)/2$. These
degeneracies implicate that the $D=2$ and $D=3$ bound - state energies
generate the ladder of the excited state energies for any given $k$ ( nodal
zeros in the wave function) and non - zero $l$ from the $l=0$ result, with
that $k$ ( for more details on this issue the reader may refer to [6-9]).
Therefore, the $D=2$ and $D=3$ bound state energies of (1) should be
correctly prescribed.

Nevertheless, the concept of a confined quantum system has originated with a
model suggested by Michels et al. [10], who proposed the idea of simulating
the effect of pressure on an atom by enclosing it in an impenetrable
spherical box. Ever since, quantum systems enclosed in boxes have received
considerable attention. An exhaustive list of publications was given by
Fr\"oman et al. [11] and more recently by Dutt et al. [12] and Varshni
[2,13-15]. It is therefore interesting to carry out systematic studies of
the effect of spherical ( D=3 ) or cylindrical ( D=2 ) impenetrable walls on
the energy levels of the singular even - power potential in (1).

Very recently, Dong and Ma [5] have used the well known elementary ansatz%
\newline
\begin{equation}
R_{0,L}(q)=exp\left( \frac{1}{2}\alpha _{0,L}~q^{2}+\frac{1}{2}\beta
_{0,L}~q^{-2}+\delta _{0,L}~ln~q\right) ,
\end{equation}
\newline
for the ground state ($k=0$) wave function (unconfined) and obtained ( for $%
D=2$) a quasi - exact, no doubt, ground state solution ( satisfying the
boundary conditions $R_{k,L}(0)=R_{k,L}(\infty )=0$. Trying to generalize
this ansatz for the first excited state, they have used\newline
\begin{equation}
R_{1,L}(q)=(a_{1,L,1}+a_{1,L,2}~q^{2}+a_{1,L,3}~q^{-2})e^{(\frac{1}{2}\alpha
_{1,L}~q^{2}+\frac{1}{2}\beta _{1,L}~q^{-2}+\delta _{1,L}~ln~q)}.
\end{equation}
\newline
When substituting (4) in (2) a set of relations, among the parameters
involved, is obtained (see equations (14a) to (14e) in [5]). A
straightforward manipulation of these would obviously lead to $%
a_{1,L,1}=a_{1,L,2}=a_{1,L,3}=0$ and a trivial solution $R_{1,L}(q)=0$ is
clearly manifested. To avoid this catastrophe the authors have imposed a
condition that the magnetic quantum number $m$ ( $l=|m|$ for $D=2$ in L
here) of the first excited state is the same as that of the ground state. At
this point, a vital clarification is in order. Although the above condition
is exotic, only the ratio $a_{1,L,2}/a_{1,L,3}=-\sqrt{a/c}$ is strictly
determined. Yet within their proposal in (4) and with $a_{1,L,1}=0$,
determining the ratio $a_{1,L,2}/a_{1,L,3}$ ought to suffice. In fact, the
ansatz suggested in (4) contradicts with the well known generalized one [4]
( used, implicitly, latter by Dong in [6])\newline
\begin{equation}
R_{k,L}(q)=F_{k,L}(q)~~exp\left( \frac{1}{2}\alpha _{k,L}~q^{2}+\frac{1}{2}%
\beta _{k,L}~q^{-2}+\delta _{k,L}ln~q\right) ,
\end{equation}
\newline
\begin{equation}
F_{k,L}(q)=q^{2k}+\sum_{i=0}^{k-1}a_{k,L,i+1}~~q^{2i},
\end{equation}
\newline
for the central singular even - power potential in (1). Which, in turn, when
substituted in (2) ( for any $k$ and $L$) implies\newline
\begin{equation}
\alpha _{k,L}=-\sqrt{a},~~\beta _{k,L}=-\sqrt{c},~~\delta _{k,L}=(\frac{3}{2}%
+\frac{b}{2\sqrt{c}}),
\end{equation}
\newline
\begin{equation}
E_{k,L}=-\alpha _{k,L}(1+4k+2\delta _{k,L}).
\end{equation}
\newline
And for $k=1$, it yields\newline
\begin{equation}
F_{1,L}(q)=q^{2}+a_{1,L,1},
\end{equation}
\newline
\begin{equation}
a_{1,L,1}=\frac{16c^{2}}{4L(L+1)c^{3/2}+8c^{2}\sqrt{a}-b^{2}\sqrt{c}%
-4cb-3c^{3/2}}
\end{equation}
\newline
with the constraint between the potential parameters\newline
\begin{equation}
4\alpha _{1,L}a_{1,L,1}=\delta _{1,1}^{2}+3\delta _{1,1}+2-L(L+1)-2\beta
_{1,1}\alpha _{1,1}.
\end{equation}
\newline
Moreover, for $k=2$ it reads\newline
\begin{equation}
F_{2,L}(q)=q^{4}+a_{2,L,2}~q^{2}+a_{2,L,1},
\end{equation}
\newline
\begin{equation}
a_{2,L,1}=\frac{16c^{2}a_{2,L,2}}{4L(L+1)c^{3/2}+8c^{2}\sqrt{a}-b^{2}\sqrt{c}%
-4cb-3c^{3/2}}
\end{equation}
\newline
\begin{equation}
G=\frac{16c^{2}}{4L(L+1)c^{3/2}+8c^{2}\sqrt{a}-b^{2}\sqrt{c}-4cb-3c^{3/2}}
\end{equation}
\newline
\begin{equation}
a_{2,L,2}=\frac{-32c^{2}}{b^{2}\sqrt{c}+12cb-8c^{2}\sqrt{a}+(35+32G\sqrt{a}%
-4L(L+1))c^{3/2}},
\end{equation}
\newline
with the constraint between the potential parameters\newline
\begin{equation}
4\sqrt{a}~a_{2,L,2}=-\delta _{2,L}^{2}-7\delta _{2,L}-12+L(L+1)+2\sqrt{ac}
\end{equation}
\newline
and so on.

Next, when the system - obeying potential (1) is confined to an impenetrable
cylindrical (D=2) or spherical (D=3) box of radius $R$, where $q=0$ locates
the center of the box, one would amend the generalized ansatz in (5) and (6)
to\newline
\begin{equation}
R_{k,L}(q)=(R^2 - q^2)~F_{k,L}(q)~~ exp\left(\frac{1}{2}\alpha_{k,L}~ q^2 +%
\frac{1}{2} \beta_{k,L}~ q^{-2} +\delta_{k,L} ln~ q \right).
\end{equation}
\newline
Where the factor $(R^2-q^2)$ ensures that $R_{k,L}(q)$ vanishes at $q=0$ and 
$q=R$ [2] and guarantees its normalizability. Maintaining the well - behaved
nature of the wave function we obtain\newline
\begin{equation}
\alpha_{k,L}=\mp\sqrt{a},~~\beta_{k,L}=-\sqrt{c},~~ \delta_{k,L}=(\frac{3}{2}%
+\frac{b}{2\sqrt{c}}),
\end{equation}
\newline
\begin{equation}
E_{k,L}=-\alpha_{k,L}(5+4k+2\delta_{k,L}),
\end{equation}
\newline
for all $k$ and $L$, where both signs of $\alpha_{k,L}$ are now admissible.

For $k=0$ and $\alpha _{0,L}=\pm \sqrt{a}$ equation (2), along with (17),
yields\newline
\begin{equation}
R^{2}=\frac{-16c^{2}}{4L(L+1)c^{3/2}-8c^{2}\alpha _{0,L}-b^{2}\sqrt{c}%
-4cb-3c^{3/2}}
\end{equation}
\newline
with the constraint\newline
\begin{equation}
\alpha _{0,L}=\frac{-b^{2}\sqrt{c}-12cb-35c^{3/2}+4L(L+1)c^{3/2}}{%
8(2R^{2}c^{3/2}+c^{2})}.
\end{equation}
\newline
At this point we should report that our results reproduce those obtained by
Varshni [2] for $D=3$ when potential (1) is confined to an impenetrable
spherical box of radius $R$.

For $k=1$ and $\alpha _{1,L}=\pm \sqrt{a}$ one obtains\newline
\begin{equation}
a_{1,L,1}=\frac{16c^{2}R^{2}}{16c^{2}+R^{2}[-8c^{2}\alpha
_{1,L}-3c^{3/2}-4cb-b^{2}\sqrt{c}+4L(L+1)c^{3/2}]},
\end{equation}
\newline
\begin{equation}
R^{2}=\frac{16a_{1,L,1}c^{3/2}\alpha
_{1,L}-99c^{3/2}-20cb+4L(L+1)c^{3/2}-8c^{2}\alpha _{1,L}-b^{2}\sqrt{c}}{%
16c^{3/2}\alpha _{1,L}}.
\end{equation}
\newline
with the constraint\newline
\begin{eqnarray}
\alpha _{1,L} &=&\frac{1}{8}\left[ -32c^{2}+a_{1,L,1}(-35c^{3/2}-12cb-b^{2}%
\sqrt{c}+4L(L+1)c^{3/2})\right.   \nonumber \\
&&\left. ~~~~+R^{2}(35c^{3/2}+12cb+b^{2}\sqrt{c}-4L(L+1)c^{3/2})\right]  
\nonumber \\
&&~~~~\times \left[ a_{1,L,1}(c^{2}+4R^{2}c^{3/2})-c^{2}\right] ^{-1},
\end{eqnarray}
\newline
In a straightforward manner, one can carry out the quasi - exact solutions
for any $k$ and dimesionality $D$ ( hence, any L) for the quantum system
defined in (1) and (2), with and without impenetrable boxes.

Nevertheless, it should be noted that the ansatz in (17) ( for the confined
system) for any $k$ addresses, with proper amendments, the $k+1$ quasi -
exact solution for the unconfined system. For example, the $k=0$ solution of
the confined system gives the $k=1$ solution for the unconfined one when $R^2
$ ( $R$ is the radius of the spherical or cylindrical box) is set equal to $%
-a_{1,L,1}$, provided that $\alpha_{k,L}=-\sqrt{a}$ to maintain the well
behaved nature of the wavefunction at $q=\infty$. Moreover, the D -
dimensional quasi - exact solutions reported above reproduce the well known
ones for $D=3$ ($L=l$) [2, and references therein] and $D=2$ ($L=l-1/2$)
[5,6, and references therein].

Of course, only a finite portion of the energy spectrum and associated
eigenfunctions are found exactly, in a closed form, for the quantum
mechanical quasi - exact solvable potentials, like the one in (1) ( a well
known characteristic of quasi - exact soluble potentials [1-6, 11-19]).
However, one may add to this that if $N$ is the number of sets of the
potential parameters, satisfying all conditions, for a given state $k$, one
could encounter sever reduction in $N$ for $k+1$ state. For example, in
table 1 of [2], Varshni obtained ( for $D=3$) 8 sets for $a=c=1$, $k=0$ and $%
\alpha_{0,L}=-\sqrt{a}$ for the confined potential (1). Whereas, for the
same parameters with $k+1$, only two sets ( for $l=0$ and $b=2.265309$ and $%
b=-14.265309$) could ${\em successfully}$ be obtained, in the present work.
And for $l=1,2,3$ none could be found.

To sum up, we have reemphasized the mixed mathematical and Physical role of
the dimensionality $D$, through the D - dimensional generalization of an
elementary ansatz in (5) and (17), and defined the related D - dimensional
quasi - exact ( conditionally - exact, in view of Znojil [19]) solutions for
the potential model in (1). Although only conditionally - exact solutions
are reported here, they remain the benchmarks for testing perturbative and
nonperturbative approximation methods for solving Schr\"odinger equation.

\end{document}